\documentclass[aps,prc,twocolumn,superscriptaddress,showpacs]{revtex4}
\usepackage{graphicx}
\usepackage{dcolumn}
\usepackage{bm}
\usepackage{slantsc}%
\usepackage{subfigure}
\usepackage{amssymb}
\newcommand{\beq}{\begin{equation}}
\newcommand{\eeq}{\end{equation}}
\newcommand{\beqar}{\begin{eqnarray}}
\newcommand{\eeqar}{\end{eqnarray}}
\newcommand{\ds}{\displaystyle}
\begin{document}

\title{ Hexagonal flow $v_6$ as a superposition of elliptic $v_2$
and triangular $v_3$ flows }

\author{L.~Bravina}
\affiliation{
Department of Physics, University of Oslo, PB 1048 Blindern,
N-0316 Oslo, Norway
\vspace*{1ex}}
\author{B.H.~Brusheim~Johansson}
\altaffiliation[Also at ]{
Oslo and Akershus University College of Applied Sciences (HIOA), 
Oslo, Norway
\vspace*{1ex}}
\affiliation{
Department of Physics, University of Oslo, PB 1048 Blindern,
N-0316 Oslo, Norway
\vspace*{1ex}}
\author{G.~Eyyubova}
\affiliation{
Skobeltsyn Institute of Nuclear Physics,
Moscow State University, RU-119991 Moscow, Russia
\vspace*{1ex}}
\author{V.L.~Korotkikh}
\affiliation{
Skobeltsyn Institute of Nuclear Physics,
Moscow State University, RU-119991 Moscow, Russia
\vspace*{1ex}}
\author{I.P.~Lokhtin}
\affiliation{
Skobeltsyn Institute of Nuclear Physics,
Moscow State University, RU-119991 Moscow, Russia
\vspace*{1ex}}
\author{L.V.~Malinina}
\altaffiliation[Also at ]{
Joint Institute for Nuclear Researches, RU-141980 Dubna, Russia
\vspace*{1ex}}
\affiliation{
Skobeltsyn Institute of Nuclear Physics,
Moscow State University, RU-119991 Moscow, Russia
\vspace*{1ex}}
\author{S.V.~Petrushanko}
\affiliation{
Skobeltsyn Institute of Nuclear Physics,
Moscow State University, RU-119991 Moscow, Russia
\vspace*{1ex}}
\author{A.M.~Snigirev}
\affiliation{
Skobeltsyn Institute of Nuclear Physics,
Moscow State University, RU-119991 Moscow, Russia
\vspace*{1ex}}
\author{E.E.~Zabrodin}
\altaffiliation[Also at ]{
Skobeltsyn Institute of Nuclear Physics,
Moscow State University, RU-119991 Moscow, Russia
\vspace*{1ex}}
\affiliation{
Department of Physics, University of Oslo, PB 1048 Blindern,
N-0316 Oslo, Norway
\vspace*{1ex}}

\date{\today}

\begin{abstract}
Partial contributions of elliptic $v_2$ and triangular $v_3$ flows
to the hexagonal $v_6$ flow are studied within the \textsc{hydjet++}
model for Pb+Pb collisions at $\sqrt{s}=2.76${\it A}~TeV. Scaling of
the ratio $v_6^{1/6}\{\Psi_2\} / v_2^{1/2}\{\Psi_2\}$ in the elliptic
flow plane, $\Psi_2$, is predicted in the range $1 \leq p_t \leq 4$\, 
GeV/$c$ for semicentral and semiperipheral collisions. Jets increase 
this ratio by about 10\% and also cause its rise at $p_t \geq 3.5$\,
GeV/$c$. The part of $v_6$ coming from $v_2$ is instantly increasing 
as the reaction becomes more peripheral, whereas the contribution of 
$v_3$ to $v_6$ drops. This behavior explains the experimentally 
observed increase of correlations between second and sixth harmonics 
and the decrease of correlations between third and sixth harmonics 
with rising impact parameter $b$. Our study favors the idea that basic 
features of the hexagonal flow can be understood in terms of the 
interplay of elliptic and triangular flows. 
\end{abstract}

%
%
\maketitle
\section{Introduction}
\label{sec1}

One of the main goals of heavy-ion experiments at ultrarelativistic 
energies is the study of properties of a new state of matter, 
quark-gluon plasma (QGP). Collider experiments with gold-gold 
collisions at $\sqrt{s} = 200$\,GeV at the Relativistic Heavy Ion 
Collider (RHIC) provided a lot of evidence that a hot and dense 
substance formed at the very beginning of the collision could be 
treated as a nearly perfect fluid \cite{rhic_exp}. Therefore, the 
whole paradigm has been changed. The plasma is no longer believed to 
be an ideal gas of noninteracting (or weakly interacting) partons, 
but rather a strongly interacting liquid \cite{shur_04}. It 
demonstrates a strong degree of collectivity, and the transverse flow 
of hadrons, particularly \textit{elliptic} flow \cite{Olli92}, is a 
very important signal that supports the hydrodynamic description of 
heavy-ion collisions. Hydrodynamic models, however, overestimate the 
flow at $p_t \geq 2$\,GeV/$c$ \cite{VPS10}, whereas conventional 
microscopic transport models usually undermine the strength of elliptic 
flow either at midrapidity \cite{BlSt02} or at high transverse momenta 
\cite{plb01,Blei05} at energies of RHIC or higher. The best description 
of the flow signal is obtained, therefore, in hybrid models, such as
\textsc{vishnu} \cite{vishnu} and \textsc{music} \cite{music}, which 
couple hydrodynamic treatment of the early stage of the expansion to 
hadron cascade model as an afterburner.

At present, the flow analysis is based on a Fourier decomposition of
the azimuthal distribution of hadrons \cite{VoZh96,PoVo98},
\beq
\ds
E \frac{d^3 N}{d^3 p} = \frac{1}{\pi} \frac{d^2 N}{dp_t^2 dy} \left[
1 + 2 \sum \limits_{n=1}^{\infty} v_n \cos{n (\phi - \Psi_n)} \right] 
\ ,
\label{eq1}
\eeq
where $\phi$ is the azimuthal angle between the transverse momentum of 
the particle and the participant event plane, each having its own
azimuth $\Psi_n$, and $p_t$ and $y$ are the transverse momentum and the 
rapidity, respectively. The flow harmonic coefficients 
\beq
\ds
v_n = \langle \cos{n (\phi - \Psi_n)} \rangle
\label{eq2}
\eeq
are obtained by averaging over all events and all particles in each
event. The first two harmonics, dubbed \textit{directed}, $v_1$, and
\textit{elliptic}, $v_2$, flow have been studied rather intensively 
during the past 15 years \cite{VPS10}, whereas the systematic study of 
higher harmonics, namely, \textit{triangular}, $v_3$, 
\textit{quadrangular} (or \textit{hexadecapole}), $v_4$, 
\textit{pentagonal}, $v_5$, and \textit{hexagonal}, $v_6$, flow began 
quite recently in the Large Hadron Collider (LHC) era \cite{qm11_flow}.
  
It is generally assumed that, in the case of noncentral collision of
two similar nuclei, remnants of the interacting nuclei fly away 
quickly, thus giving space for expansion of the overlapped volume. 
In the transverse plane this area resembles an ellipse; therefore, odd 
harmonics of anisotropic flow, such as $v_3,\ v_5,$ etc., can be 
neglected because of the symmetry considerations. The concept of 
participant triangularity due to initial-state fluctuations was first 
introduced in \cite{AlRo10}. In model simulations, the triangular flow 
signal was found to be directly proportional to the participant 
triangularity. After that, correlations were studied between the 
higher-order harmonic eccentricity coefficients $\varepsilon_n$, linked 
to participant plane angles $\Phi_n$, and the final anisotropic flow 
coefficients $v_n$ and their final anisotropic flow angles $\Psi_n$;
see, e.g., \cite{AGLO10,QPBM10,GGLO12,QH11,TY12}. This analysis was
done within both ideal and viscous relativistic hydrodynamics with 
Monte Carlo$-$Glauber or color glass condensate (CGC) initial 
conditions. One of the interesting observations is that just the first
few flow harmonics survive after the hydrodynamic evolution despite 
the fact that the initial spacial anisotropies are of the same order
\cite{QPBM10}. The characteristic mode mixing between the different 
order flow coefficients has been revealed \cite{QH11,TY12,GGLO12}.
It is found that the final plane angles $\Psi_n,\ n > 3$ seem to be 
uncorrelated with the corresponding participant plane angles $\Phi_n,
\ n>3$, associated with initial anisotropies \cite{HS13}. In contrast, 
the response of the elliptic flow to ellipticity, as well as that of
the triangular flow to triangularity, is approximately linear 
\cite{QH11}. Is it because of the crosstalk of several harmonics, and 
which harmonics play a major role in this process? To answer these 
questions it would be important to study the influence of $v_2$ and 
$v_3$, linked to elliptic and triangular anisotropies, respectively, 
on higher harmonics of the anisotropic flow. For our analysis the 
\textsc{hydjet++} model \cite{hydjet++} is employed. The basic 
principles of the model are given in Sec.~\ref{sec2}.  

\section{Model}
\label{sec2}

The Monte Carlo event generator \textsc{hydjet++} is a superposition 
of two event generators, \textsc{fastmc} \cite{fastmc1,fastmc2} and 
\textsc{pyquen} \cite{hydjet}, describing soft and hard parts of 
particle spectra in ultrarelativistic heavy-ion collisions at energies 
from RHIC ($\sqrt{s} = 200$\,GeV) to LHC ($\sqrt{s} = 5.5$\,TeV). Both 
\textsc{fastmc} and \textsc{pyquen} generate particles independently. 
Their partial contributions to the total event multiplicity depend on 
collision energy and centrality. The soft part of a single event in 
\textsc{hydjet++} is a thermal hadronic state treated within the 
framework of parametrized hydrodynamics \cite{fastmc1,fastmc2}. The 
hard part of this event is represented by a multiple scattering of 
hard partons in a hot and dense medium, such as quark-gluon plasma. It 
accounts for the radiative and collisional energy losses \cite{pyquen} 
and shadowing effect \cite{shad}. Further details of the 
\textsc{hydjet++} model can be found elsewhere 
\cite{hydjet++,fastmc1,fastmc2,hydjet}. Below we concentrate on the 
simulation of anisotropic flow in the recent version of 
\textsc{hydjet++} \cite{hydjet_13}.
 
In the case of noncentral collisions of identical nuclei the overlap
area has a characteristic almond shape. This ellipsoid posseses the 
initial coordinate anisotropy, which is a function of impact parameter 
$b$ and nuclear radius $R_A$, $\epsilon_0(b) = b/(2 R_A)$. In the 
azimuthal plane the transverse radius of the fireball reads 
\cite{fastmc2}
\beq \ds
R_{\rm ell}(b, \phi) = R_{\rm fo}(b) \frac{\sqrt{1 - \epsilon^2(b)}}
{\sqrt{1 + \epsilon(b) \cos{2\phi}}}\ .
\label{r_ell}
\eeq
Here $\phi$ is the azimuthal angle and $R_{\rm fo}(0)$ is the model
parameter that determines the scale of the fireball transverse size 
at freeze-out. The pressure gradients are stronger in the direction of 
the short axis in the transverse plane. Thus, the initial spatial 
anisotropy is transformed into the momentum anisotropy, which results 
in the anisotropy of the flow. The azimuthal angle of the fluid 
velocity vector $\phi_{\rm fl}$ is linked to the azimuthal angle $\phi$ 
via \cite{fastmc2}
\beq
\ds
\frac{\tan{\phi_{\rm fl}}}{\tan{\phi}} = \sqrt{\frac{1 - \delta(b)}
{1 + \delta(b)}}\ ,
\label{phi}
\eeq
with $\delta(b)$ being the flow anisotropy parameter. In the employed
version of \textsc{hydjet++} both spatial and flow anisotropies, 
$\epsilon(b)$ and $\delta(b)$, are proportional to the initial spatial
anisotropy $\epsilon_0 = b/(2 R_A)$.   

To introduce the triangular flow the transverse radius of the freeze-out 
surface is modified further [cf. Eq.~(10) from \cite{AGLO10}]:
\beq
\ds
R (b,\phi) = R_{ell}(b,\phi) \{ 1 + \epsilon_3(b) \cos{\left[
3(\phi - \Psi_3)\right]} \} \ ,
\label{r_3}
\eeq
where the new phase $\Psi_3$ is randomly distributed with respect to 
the position of the reaction plane $\Psi_2$. It means, in particular, 
that the integrated triangular flow measured in the $\Psi_2$ plane is 
zero, in accordance with the experimental observations. Similarly to 
$\epsilon(b)$, the new parameter $\epsilon_3(b)$, which is responsible 
for emergence of the triangular anisotropy, can be either linked to 
initial eccentricity $\epsilon_0(b)$ or treated as a free parameter.
   
It is worth mentioning here several important points. Like many other
hydrodynamic models, \textsc{hydjet++} does not consider directed 
flow; i.e., $v_1$ of particles is essentially zero. The model describes 
the midrapidity area of heavy-ion collisions rather than the 
fragmentation ones. Recent measurements of the directed flow of charged 
particles done by the ALICE Collaboration at midrapidity in lead-lead 
collisions at $\sqrt{s} = 2.76$\,TeV \cite{v1_alice} show that $v_1$ is 
order(s) of magnitude weaker than $v_2$ and $v_3$. Then, in contrast to 
event-by-event (EbE) hydrodynamics, \textsc{hydjet++} has no evolution 
stage and, therefore, cannot trace, e.g., propagation of energy and 
density fluctuations of the initial state, the so-called hot spots. It 
deals already with the final components of anisotropic flow. Lacking 
the EbE fluctuations, the model-generated ratios of different flow 
harmonics could be directly confronted only with the ratios obtained 
from EbE analysis of the data. This is not the case, however, because 
the data on flow harmonics are averaged over the whole statistics before 
performing the analysis of ratios, such as $v_n^{1/n} / v_2^{1/2}$. It 
leads to acquiring an extra multiplier to which the model results (or 
data) should be adjusted; see \cite{GO10} for details.

The elliptic flow of particles contributes to all even harmonics, i.e.,
$v_4,\ v_6,$ etc. For instance, quadrangular flow in \textsc{hydjet++} 
is determined by the elliptic flow of particles, governed by 
hydrodynamics, and particles coming from jets \cite{prc13,sqm11}. The 
interplay between the elliptic and triangular flows will result in the
appearance of odd higher harmonics in the model. Similarly to $v_2$, 
triangular flow should contribute separately to $v_6$, $v_9$, etc. The 
goal of our study of the hexagonal flow, $v_6$, is, therefore, twofold. 
First, the partial contributions of $v_2$ and $v_3$, each having its 
own flow angle $\Psi_2$ and $\Psi_3$, to $v_6$ should be analyzed. Of 
particular interest are the features of the distributions 
$v_6\{\Psi_2\} (p_t)$ and $v_6\{\Psi_3\} (p_t)$. Second, the model
allows one to investigate the influence of nonflow correlations,
arising from jet fragmentation and resonance decays, on the flow 
harmonics. The previous study \cite{prc13,sqm11} of the $v_4/v_2^2$ 
ratio revealed that the jet contribution to this ratio is quite
substantial compared to the slight modification caused by the decays 
of resonances. But, before the analysis of generated spectra, we have 
to estimate individual contributions of elliptic and triangular flow 
to $v_6$ within the framework of relativistic ideal hydrodynamics.
  
\section{$v_6$ as a function of $v_2$ and $v_3$}
\label{sec3}

As was shown within the approach suggested in \cite{BO06}, the 
freeze-out distribution of fast particles obtained by a saddle-point 
integration is proportional to the exponential
\beq \ds
\frac{d^3 N}{dy d^2 p_t} \propto \exp{ \left(\frac{p_t u_{\rm max} - 
m_t u_{\rm max}^0}{T} \right)}\ ,
\label{expon}
\eeq
where $u = (u^0, u_\parallel, u_\perp)$ is the fluid 4-velocity,
$u_\parallel \equiv u_{\rm max},\ v_{\rm max} = u_{\rm max}^0/
u_{\rm max}$, $y$ is the rapidity, $T$ is the temperature and $m_t = 
\sqrt{m^2 + p_t^2}$ is the transverse mass of a particle. The method 
utilizes the fact that fast particles come from regions of the 
freeze-out hypersurface where the $u_\parallel$, which is parallel to 
the particle's transverse momentum $\overrightarrow{p_t}$, is close to 
its maximum value $u_{\rm max}$ \cite{BO06}. Assuming for the sake of 
simplicity a single event plane and expanding $u_{\rm max}(\phi)$
in Fourier series, one gets
\beq \ds
u_{\rm max}(\phi) = u_{\rm max} \left[ 1 + 2 \sum 
\limits_{n=1}^{\infty} V_n \cos{(n \phi)} \right] \ .
\label{u_exp}
\eeq
Denoting 
$$\ds a = \frac{p_t - m_t v_{\rm max}}{T} \,\, u_{\rm max}\ ,$$ 
we obtain from Eqs.~(\ref{eq1}), (\ref{expon}), and (\ref{u_exp})
\beq \ds
\exp{ \left\{ a \left[ 1 + 2 \sum \limits_{n=1}^{\infty} V_n 
\cos{(n \phi)} \right] \right\} } =
1 + 2 \sum \limits_{n=1}^{\infty} v_n \cos{(n \phi)} \ .
\label{exp_flow}
\eeq
Then, the expressions for the elliptic and triangular flows read
\beqar \ds
v_2 = a V_2 & \equiv & \frac{p_t - m_t v_{\rm max}}{T}\, 
u_{\rm max} V_2\ , \\
v_3 = a V_3 & \equiv & \frac{p_t - m_t v_{\rm max}}{T}\, 
u_{\rm max} V_3\ ,
\label{v2_v3}
\eeqar
respectively. It is easy to see that the quadrangular flow depends on
both $V_2$ and $V_4$:
\beq \ds
v_4 = \frac{1}{2} a^2 V_2^2 + aV_4\ .
\label{v4}
\eeq
Since the last term in Eq.(\ref{v4}) 
$a V_4 \ll a^2 V_2^2 \equiv v_2^2$ at $p_t \rightarrow \infty$, we 
regain the familiar result $\ds v_4 \cong \frac{1}{2} v_2^2$ \cite{BO06}. 
For the hexagonal flow one gets, after the straightforward calculations,
\beq \ds
v_6 = \frac{1}{6} (a^2 V_2)^3 + \frac{1}{2} (a V_3)^2 + a V_6 + 
3 (a V_2) (a V_4) \ .
\label{v6_tot}
\eeq
Taking into account that at high transverse momenta $a V_4 \ll v_2^2$
and $a V_6 \ll a^2 V_3^2 \equiv v_3^2$ we arrive at the simple expression
\beq \ds
v_6 \cong \frac{1}{6} v_2^3 + \frac{1}{2} v_3^2 \ .
\label{v6}
\eeq 

Note again that this result was obtained under the assumption of a 
single event plane. We have learned in the past few years, however,
that each of the flow harmonics $v_n$ possesses its own event plane 
$\Psi_n$ not necessarily coinciding with the others. The interplay 
between different event planes can be very important, and one should 
consider Eq.~(\ref{v6}) as a first-order approximation. Model results 
for the hexagonal flow and its correlations with the elliptic and 
triangular flows are given below.  

\section{Results and discussion}
\label{sec4}
   
\begin{figure}
\resizebox{\linewidth}{!}{
\includegraphics[scale=0.60]{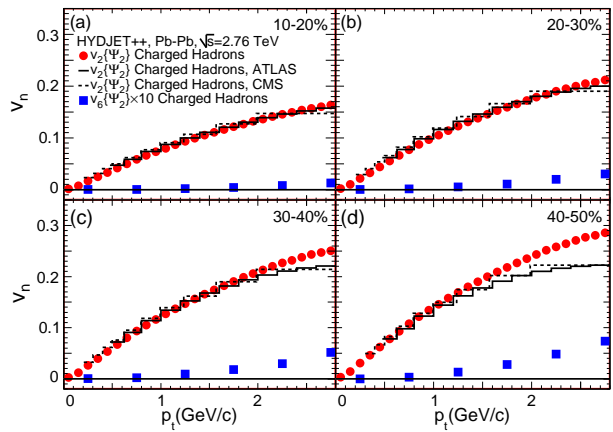}
}
\caption{(Color online)
Transverse momentum dependencies of $v_2\{\Psi_2\}$ (circles) and 
$v_6\{\Psi_2\}$ (squares) of charged hadrons calculated within the 
\textsc{hydjet++} for Pb+Pb collisions at $\sqrt{s} = 2.76$\,TeV 
at centralities (a) $\sigma/\sigma_{\rm geo} = 10 - 20\%$, (b)
$20 - 30\%$, (c) $30 - 40\%$, and (d) $40 - 50\%$. Solid and dashed
histograms show experimental data on $v_2$ taken from ATLAS 
\cite{atlas_12} and CMS \cite{cms_13}, respectively.
\label{fig1} }
\end{figure}

\begin{figure}
\resizebox{\linewidth}{!}{
\includegraphics[scale=0.60]{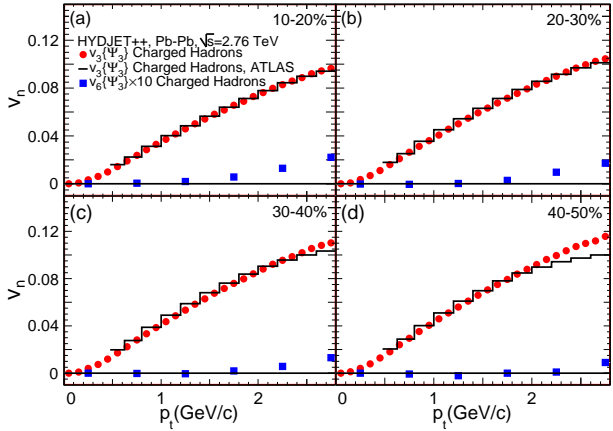}
}
\caption{(Color online)
The same as Fig.~\protect\ref{fig1} but for triangular and hexagonal
flows in the $\Psi_3$ plane. Histograms show experimental data on 
$v_3$ taken from \cite{atlas_12}.
\label{fig2} }
\end{figure}

To study the formation of the hexagonal flow in the model ca. 
$2 \times 10^6$ lead-lead collisions were generated for each of four 
centralities $\sigma/\sigma_{\rm geo} = 10 - 20$\%, $20 - 30$\%, 
$30 - 40$\%, and $40 - 50$\%. Transverse momentum distributions of 
$v_6$ in $\Psi_2$ and $\Psi_3$ planes are shown in Figs.~\ref{fig1} 
and \ref{fig2}, respectively, together with the corresponding 
distributions for the elliptic and triangular flows. Available 
experimental data for $v_2(p_t)$ and $v_3(p_t)$ are plotted onto the 
model calculations as well. The agreement with the data for both flow 
harmonics is fair. A detailed comparison of the model results with the 
data is given in \cite{hydjet_13}. Recall, that in contrast to many 
other hydrodynamic models the \textsc{hydjet++} model demonstrates a 
drop of elliptic flow at $p_t \geq 3$\,GeV/$c$ \cite{prc09,sqm09}. This 
drop is attributed in the model to the interplay of soft hydrodynamic 
processes and hard jets. In ideal hydrodynamics, particles with higher 
transverse momenta are carrying larger elliptic flow. However, the 
number of these particles decreases exponentially with rising $p_t$, 
and after certain $p_t$ the particle spectrum is dominated by hadrons 
coming out from quenched jets. The elliptic flow of the jet hadrons is 
much lesser than the flow of hydro-induced hadrons; thus, the resulting 
flow of high-$p_t$ particles drops (to almost zero modulo path-length 
dependence of in-medium partonic energy loss).

It appears that the hexagonal flow in \textsc{hydjet++} is weak but
not zero in both $\Psi_2$ and $\Psi_3$ planes. In the $\Psi_2$ plane
it starts to rise at $p_t \geq 1.5$\,GeV/$c$ in semiperipheral 
collisions with $\sigma/\sigma_{\rm geo} \geq 30\%$. Here we observe a 
clear tendency that $v_6$ of charged hadrons with high transverse 
momenta increases with rising impact parameter. In the $\Psi_3$ plane 
the high-$p_t$ tail of the distribution is presented as well. The 
generated $v_6\{\Psi_3\}(p_t)$ seems to become a bit weaker at $1 \leq 
p_t \leq 2.5$\,GeV/$c$ with increasing $b$, despite the fact that 
triangular flow slightly increases. This peculiarity is clarified in 
our study below.

\begin{figure}
\resizebox{\linewidth}{!}{
\includegraphics[scale=0.60]{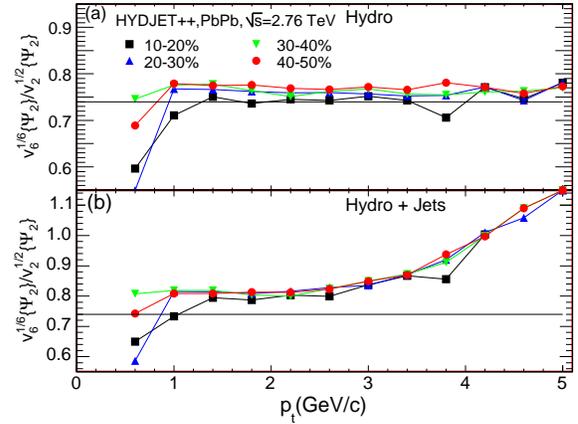}
}
\caption{(Color online)
Ratio $v_6^{1/6}/v_2^{1/2}$ as a function of $p_t$ in the $\Psi_2$
event plane for charged particles, originated from (a) soft processes
only and (b) both soft and hard processes, in \textsc{hydjet++}
simulations of Pb+Pb collisions at $\sqrt{s} = 2.76$\,TeV. The
reaction centralities are $10 - 20$\% (squares), $20 - 30$\% 
(triangles pointing up), $20 - 30$\% (triangles pointing down), and 
$40 - 50$\% (circles). Solid lines in both plots show the prediction 
of ideal hydrodynamics for this ratio at high $p_t$, namely, 
$v_6^{1/6}/ v_2^{1/2} = (1/6)^{1/6} \approx 0.74$.
\label{fig3} }
\end{figure}

\begin{figure}
\resizebox{\linewidth}{!}{
\includegraphics[scale=0.60]{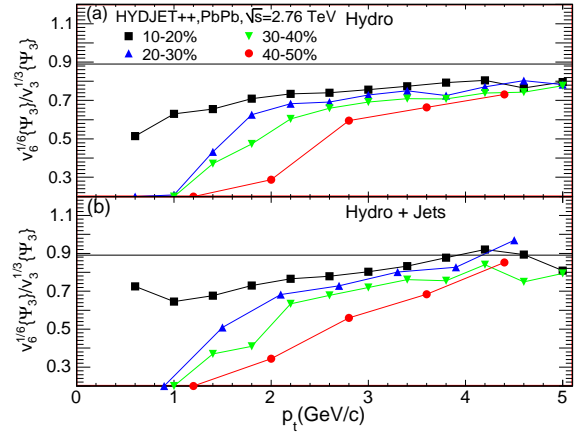}
}
\caption{(Color online)
The same as Fig.~\protect\ref{fig3} but for the ratio $v_6^{1/6}/
v_3^{1/3}$ vs. $p_t$ in the $\Psi_3$ event plane. Solid lines in both 
plots show the prediction of ideal hydrodynamics for this ratio at 
high $p_t$, namely, $v_6^{1/6}/v_3^{1/3} = (1/2)^{1/6} \approx 0.89$.
\label{fig4} }
\end{figure}

To check the scaling trends in the behavior of different flow harmonics 
the ratio $v_n^{1/n} / v_2^{1/2}$ is employed. The ratio $v_6^{1/6}
(p_t) / v_2^{1/2}(p_t)$ in the $\Psi_2$ plane is displayed in 
Fig.~\ref{fig3} (a) for hadrons participated only in the hydrodynamic 
process and in Fig.~\ref{fig3}(b) for all hadrons in the system. Note 
that the hexagonal flow here is determined with respect to $\Psi_2$ 
plane and not its own $\Psi_6$ plane. One can see the real scaling at 
$p_t \geq 1$\,GeV/$c$, where all curves are on top of each other. For 
``hydrodynamic" particles the relation $v_6/v_2^3 \approx 1/6$ is 
fulfilled with good accuracy already at $p_t = 1$\,GeV/$c$. The effect 
of jets is twofold. First of all, hadrons from jets increase the 
considered ratio by $\sim 10\%$ in the interval $1 \leq p_t \leq 
3$\,GeV/$c$, as demonstrated in Fig.~\ref{fig3}(b). Second, at larger 
transverse momenta the ratio starts to rise further in contrast to the
plateau in the hydrodynamic case. 

The situation with the ratio $v_6^{1/6}(p_t) / v_3^{1/3}(p_t)$ in the
$\Psi_3$ plane, which is depicted in Fig.~\ref{fig4}, is not so clear. 
This ratio is below the ideal high-$p_t$ limit $v_6 / v_3^2 \approx 
1/2$, but steadily increases to it with rising transverse momentum. 
Jets also increase this ratio and make its rise a bit steeper. In 
contrast to the scaling in the $\Psi_2$ plane, the ratio $v_6^{1/6} / 
v_3^{1/3}$ in the $\Psi_3$ plane decreases for more peripheral 
collisions. 
   
\begin{figure}
\resizebox{\linewidth}{!}{
\includegraphics[scale=0.60]{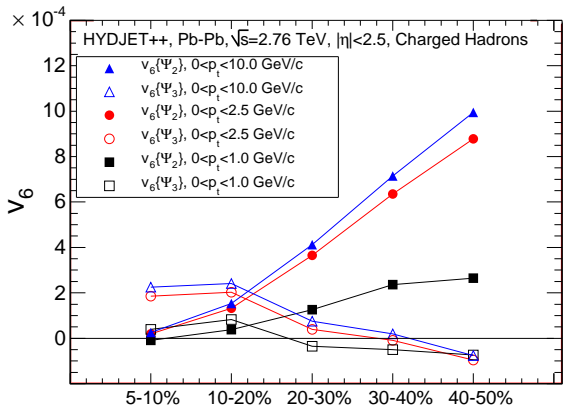}
}
\caption{(Color online)
Hexagonal flow $v_6\{\Psi_2\}$ and $v_6\{\Psi_3\}$ of charged particles 
vs. centrality in Pb+Pb collisions at $\sqrt{s} = 2.76$\,TeV
simulated by \textsc{hydjet++}. The pseudorapidity interval is 
$|\eta| < 2.5$. Solid and open symbols indicate $v_6\{\Psi_2\}$ and
$v_6\{\Psi_3\}$ of hadrons with transverse momenta below 10\,GeV/$c$
(triangles), 2.5\,GeV/$c$ (circles), and 1\,GeV/$c$ (squares), 
respectively.
\label{fig5} }
\end{figure}

This means that the partial contributions of elliptic and triangular 
flows to the projections of the hexagonal flow onto $\Psi_2$ and
$\Psi_3$ planes are changing with centrality. 
Figure~\ref{fig5} presents $v_6$, averaged in several $p_t$
intervals, as a function of centrality in both $\Psi_2$ and $\Psi_3$
planes. Although the absolute magnitude of the signals depends on the
selected $p_t$ intervals, the tendencies in the $v_6$ development
are clearly revealed. Namely, $v_6\{\Psi_2\}$ is weak in semicentral
collisions but gradually increases for more peripheral reactions.
This issue is supported by recent CMS data on hexagonal flow extracted 
by different methods \cite{cms_high_harm}.
And vice versa, $v_6\{\Psi_3\}$ is maximal in semicentral collisions 
and then drops. Summarizing information provided by Eq.~(\ref{v6}) 
and Figs.~\ref{fig1}, \ref{fig2}, and \ref{fig5}, we arrive at the
following scenario. For central topologies triangular flow is stronger
than the elliptic one; therefore, it makes the main contribution to
the hexagonal flow. The event plane $\Psi_6$ is closer to the $\Psi_3$
rather than the $\Psi_2$ one. (Recall, that there are no genuine 
hexagonal deformations in the \textsc{hydjet++} model that can account 
for the formation of genuine $v_6$.) In peripheral topologies elliptic 
flow dominates over the triangular one. Thus, the resulting hexagonal
flow event plane $\Psi_6$ should be oriented closer to $\Psi_2$.
In other words, in semicentral collisions $\Psi_6$ is more strongly 
correlated with $\Psi_3$, whereas in more peripheral collisions 
$\Psi_6$ is correlated with $\Psi_2$.  

\begin{figure}
\vspace{0.1cm}
\resizebox{\linewidth}{!}{
\includegraphics[scale=0.60]{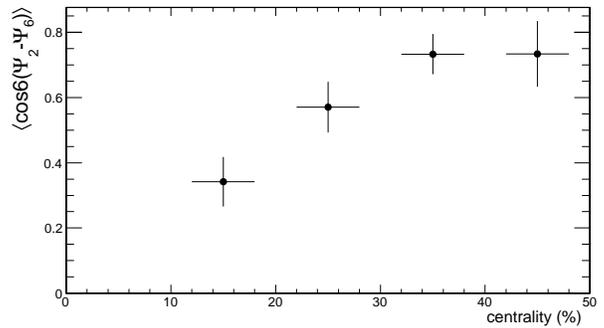}
}
\caption{
Two-plane correlator $\cos{\langle 6(\Psi_2 - \Psi_6) \rangle}$ 
as a function of centrality for charged hadrons in \textsc{hydjet++} 
simulated Pb+Pb collisions at $\sqrt{s} = 2.76$\,TeV.
\label{fig6} }
\end{figure}

\begin{figure}
\resizebox{\linewidth}{!}{
\includegraphics[scale=0.60]{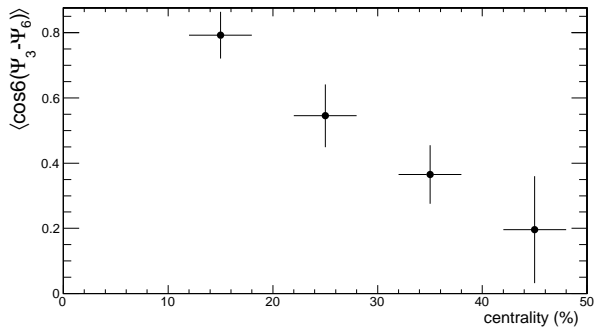}
}
\caption{
The same as Fig.~\protect\ref{fig6} but for two-plane correlator
$\cos{\langle 6(\Psi_3 - \Psi_6) \rangle}$.
\label{fig7} }
\end{figure}

To see this interplay more distinctly, we apply the method of
event plane correlators \cite{BLO11,jia_corr,BOP13}. For each flow
harmonic of $n$th order one has to determine the event flow vector
$\overrightarrow{Q_n}$ and the event plane angle $\Psi_n$ following, 
e.g., prescription of \cite{VPS10,PoVo98}
\beqar \ds
\nonumber
\overrightarrow{Q_n} &=& (Q_{n,x} , Q_{n,y}) = \left( 
\sum \limits ^{}_{i} w_i
\cos{(n \phi_i)} , \sum \limits ^{}_{i} w_i \sin{(n \phi_i)} \right)
\\
\label{q_n}
      &=& \left( Q_n \cos{(n \Psi_n)} , Q_n \sin{(n \Psi_n)} \right)
\ , 
\eeqar
\beq \ds
\tan{(n \Psi_n)} = \frac{Q_{n,y}}{Q_{n,x}}\ ,
\label{psi_n}
\eeq
where $w_i$ and $\phi_i$ are the weight and the azimuthal angle of the 
$i$th particle in the laboratory system, respectively. The correlators 
between arbitrary $l$ event planes of order $k_l$ have the form
$ \langle \cos{( \sum \limits ^{k_{l_{\rm max}}}_{k=k_{l_{\rm min}}}
k c_k \Psi_k )} \rangle $ with the constraint
$ \sum \limits ^{k_{l_{\rm max}}}_{k=k_{l_{\rm min}}} k c_k = 0 $. 
In our case of just two planes, $(\Psi_2, \Psi_6)$ and 
$(\Psi_3, \Psi_6)$, the correlators are simply 
$\langle \cos{6 (\Psi_2 - \Psi_6)} \rangle$ and 
$\langle \cos{6 (\Psi_3 - \Psi_6)} \rangle$, respectively. Both
correlators were extracted from the \textsc{hydjet++} events by the
method applied for analysis of experimental data \cite{jia_corr}. This 
approach implies separation of a single event into two forward-backward
symmetric subevents with a pseudorapidity gap in between, and takes
into account resolution corrections for each of the event planes; see
\cite{jia_corr} for details and also \cite{BOP13} for generalization 
of the method. Moreover, to avoid ambiguity in the interpretation of
the results, we artificially increased the triangularity of the
freeze-out hypersurface. The obtained correlators are displayed in
Figs.~\ref{fig6} and \ref{fig7}. In contrast to Fig.~\ref{fig5}, here 
the correlations are investigated between the different event planes 
and not between the flow harmonics projected onto $\Psi_2$ or $\Psi_3$ 
planes. We see that the correlator $\langle \cos{6 (\Psi_2 - \Psi_6)} 
\rangle$ increases for more peripheral collisions, whereas the 
correlator $\langle \cos{6 (\Psi_3 - \Psi_6)} \rangle$ drops. Similar 
centrality dependencies were observed by the ATLAS Collaboration as 
well \cite{jia_corr}. Such a behavior has a simple explanation. The 
event plane $\Psi_6$ becomes closer to the $\Psi_2$ one as the 
hexagonal flow is strongly determined by the $v_2$ for the peripheral 
collisions. Because $v_3$ is randomly oriented with respect to $v_2$, 
the correlations between the $\Psi_6$ and the $\Psi_3$ become weaker.
Recently, various two- and many-plane correlators were studied in 
\cite{BOP13} within the microscopic a multiphase transport \textsc{ampt} 
model. Very good agreement with the experiment is demonstrated. However, 
the authors attribute the drop of the correlations between the third and 
the sixth harmonics to the decrease of the triangular flow itself. This 
is not the case, because the magnitude of the $v_3$ is approximately the 
same, as one can see, e.g., in Fig.~\ref{fig2}. In our opinion, the
falloff is driven by two reasons: (i) domination of $v_2$ over $v_3$ 
in semiperipheral and peripheral collisions, and (ii) absence of
correlations between the $\Psi_2$ and $\Psi_3$.

Finally, the contribution of the genuine hexagonal fluctuations to 
the final hexagonal flow should be weak. The experimentally observed
event plane correlations and other features of $v_6$ are reproduced
in terms of interplay between the second and the third flow harmonics.  
 
\section{Conclusions}
\label{sec5}

The hexagonal flow $v_6$ is studied within the \textsc{hydjet++} 
model in Pb+Pb collisions at $\sqrt{s} = 2.76$\,TeV and centralities 
$10\% \leq \sigma/\sigma_{\rm geo}\leq 50\%$. In contrast to the 
majority of hydrodynamic models, the \textsc{hydjet++} model combines 
parametrized hydrodynamics with jets. Only second and third flow 
harmonics are generated at the freeze-out hypersurface in the present 
version of the  model; therefore, the hexagonal flow originates solely 
as a result of nonlinear hydrodynamic response, 
$v_6 \sim v_2^3 + v_3^2$. The following conclusions can be drawn.

(1) Scaling of the ratio $v_6^{1/6} \{ \Psi_2 \} / v_2^{1/2} \{ \Psi_2 
\} $ is observed in the $\Psi_2$ event plane within
the indicated centrality interval. No scaling is found for the ratio
$v_6^{1/6} \{ \Psi_3 \} / v_3^{1/3} \{ \Psi_3 \} $.

(2) Jets increase both ratios by $10\% - 15\%$ and lead to rising
high-$p_t$ tails at $p_t \geq 3$\, GeV/$c$.

(3) The behavior of the plane correlators 
$\langle \cos{6 (\Psi_2 - \Psi_6)} \rangle$  and
$\langle \cos{6 (\Psi_3 - \Psi_6)} \rangle$ is in line with the
experimental observations and with the centrality dependencies of
$v_6$ on $v_2$ and $v_3$ in the $\Psi_2$ and $\Psi_3$ event planes,
respectively. These findings strongly favor the idea that basic 
features of the hexagonal flow can be understood as a result of 
contributions of elliptic and triangular flows and their interplay.
Original hexagonal initial fluctuations seem to play a minor role
in the formation of $v_6$.   

{\it Acknowledgments.}
Fruitful discussions with L. Csernai and J.-Y. Ollitrault are 
gratefully acknowledged.
This work was supported in parts by the Department of Physics, UiO, 
Russian Foundation for Basic Research under Grant No. 12-02-91505 
and a Grant of the President of Russian Federation for Scientific 
School Supporting No. 3920.2012.2.

\end{document}